# Orbits in Large Aluminum Clusters: Five-Pointed Stars

## G. V. Shpatakovskaya

*Institute of Mathematical Modeling, Russian Academy of Sciences, Miusskaya pl. 4a, Moscow, 125047 Russia*
*e-mail: shpat@imamod.ru*
Received August 2, 2000

The distinctions in the mass spectra of large sodium $Na_N$ and aluminum $Al_N$ clusters are discussed. A semiclassical method is used to describe the shell effects within a spherical jellium model. It allows one to analyze the relative role of different classical trajectories in the formation of electronic supershells in clusters of various sizes at zero and finite temperatures. A criterion for the hardness of the self-consistent potential is formulated. The conjecture that the five-point-star trajectories make the main contribution to the spectral oscillations for large soft-potential $Al_N$ ($250 < N < 900$) clusters is substantiated. The computational results are in agreement with the mass spectra of the $Al_N$ clusters at $T \simeq 300$ K. © *2000 MAIK "Nauka/Interperiodica"*.

**1.** The oscillations in the mass spectra of metal clusters may be caused by both the shell structure of electronic spectra and the positioning of ions in lattice sites [1]. Experiments show that the oscillations in the mass spectra of large aluminum $Al_N$ [2–4] and sodium $Na_N$ [5] clusters ($N$ is the number of atoms in a cluster) differ significantly in shape. Whereas the oscillations for sodium proceed with beats, the aluminum clusters with $N > 250$ exhibit sinusoidal behavior with a frequency approximately twice that for sodium. The spectra of the $Al_N$ clusters of smaller sizes represent an intricate pattern without any distinct period. In the literature, the cause of this distinction is discussed in terms of classical trajectories of electron motion in a self-consistent potential (the number of electrons in a cluster is $N_e = wN$, where $w$ is the valence of a metal).

In [3], an attempt was undertaken to explain the experiment by invoking a spherical jellium model and the quasiclassical theory [6–9]. It was conjectured that, contrary to a hard potential of $Na_N$ clusters, in which a triangular trajectory and a square trajectory of a close frequency dominate (the oscillations with beats result from the interference of the relevant contributions), in a soft potential of $Al_N$ clusters with $250 < N < 900$, the main contribution comes from a single trajectory shaped like a five-pointed star. According to this theory, the clusters of larger size should have triangular and, later, square trajectories (this is confirmed by the self-consistent computation [2] of the density of states for $N_e = 4940$), leading to a change in the oscillation frequency.

An alternative explanation was suggested in [4], where the mass spectra of "cold" ($T = 100$ K) $Al_N$ clusters were experimentally measured and analyzed over a very wide range of $N$ values ($250 < N < 10000$). An analysis of the spectra showed that the oscillation maxima, numbered sequentially by the index $k$ ($k > 25$), appeared with a constant frequency over the entire range studied and fitted the law $N \simeq 0.0104 k^3$, which is readily explained by the atomic positioning in an octahedral lattice. Accordingly, the cluster shape is not a sphere but an octahedron, so that the shell filling corresponds to the assembling of one of its faces. Evidently, the spherical jellium model with uniformly distributed ions inside the sphere is not adequate in this case.

In [10], the assumption about the dominant contribution from a five-point-star orbit in the aluminum clusters was ruled out by the quantum-mechanical calculation of the density of states for $N_e = 1000$ electrons in the Saxon–Woods potential.

Nevertheless, the positions of the maxima observed in [2] for $250 < N < 430$ at $T = 295$ K agree well with the results of self-consistent calculations carried out in the same work with the jellium model, while the comparison of the mass spectra of $Al_N$ observed at $T = 110$ K and $T = 295$ K for $N < 250$ reveals that the temperature has an appreciable effect on the shapes of the corresponding curves, indicating that the lattice melts upon temperature elevation and manifesting the electronic shell structure.

Therefore, although the contribution of the ion lattice to the oscillations in mass spectra likely dominates in aluminum at low temperatures and, hence, the spherical jellium model is inadequate in this temperature range, the role of the lattice diminishes with increasing temperature, rendering the jellium model more applicable. It is assumed in this work that the electronic structure reveals itself upon melting the lattice of "hot" clusters and that the melting temperature can be experimentally determined from the changes in the mass spectra at $N \sim 1000$. However, whereas the jellium calculations were carried out for sodium clusters over a wide range of $N$ values and for different temperatures [11], the respective calculations for aluminum are still lacking.

In this work, this gap is filled by studying the dependence of the oscillating part of the electronic free energy of an aluminum cluster on its size and temperature, and the assumption about the dominant role of a five-point-star orbit in the oscillations of $Al_N$ spectra at $T \simeq 300$ K is confirmed for the $N$ numbers experimentally observed in [3].

**2.** The semiclassical approach [12, 13] used in this work is based on the spherical jellium model and the extended Thomas–Fermi (ETF; see references in [14]) model, whose solution, namely, the electron density $n(r)$, the chemical potential $\mu$, the self-consistent potential $U(r)$, and the corresponding electronic free energy $F(N_e,T)$, are assumed to be known. The ETF model well describes the average characteristics of a system, while the shell effects of interest will be studied using the following expression for the correction to the free energy (in atomic units) [13, 15]:

$$\Delta F_{sh} = -\int_{-\infty}^{\mu} d\mu' \frac{\hat{X}_{\mu'}(T)}{\sinh(\hat{X}_{\mu'}(T))} \Delta N_{sh}(\mu', 0). \quad (1)$$

Here, the operator $\hat{X}_\mu(T) = -i\pi T \partial/\partial\mu$ and $\Delta N_{sh}(\mu, 0)$ is the shell correction to the number of states with energies below $\mu$ without regard for explicit temperature dependence. In the semiclassical approximation

$$\Delta N_{sh}(\mu, 0) = \Delta N_{sh}(\mu)$$
$$= \frac{2}{\pi} \sum_{k,s}' \frac{(-1)^{k+s}}{k} \int_0^{\lambda_\mu} d\lambda \lambda \sin[2\pi(k\nu_\mu(\lambda) + s\lambda)], \quad (2)$$

where summation over $k$ and $s$ goes from $-\infty$ to $+\infty$; the prime over the sum sign indicates that the nonoscillating term with $k = s = 0$ is omitted; $S_{\mu\lambda} = \pi\nu_\mu(\lambda)$ is the radial action between the turning points of the electron motion with energy $\mu$ and orbital angular momentum $\lambda$ in the potential $U(r)$; $\lambda_\mu = p_\mu(r_0)r_0$ is the maximum orbital angular momentum for energy $\mu$; $r_0$ is the point at which the $p_\mu(r)r$ function is maximum; and $p_\mu(r) = \sqrt{2(\mu - U(r))}$.

In the semiclassical theory, the integral in Eq. (2) is calculated by the saddle-point method and the sum of contributions from the saddle points $\bar{\lambda}_j$ has the form

$$\Delta N_{sh}(\mu) = \sum_j \frac{2\bar{\lambda}_j}{\pi\sqrt{\nu_\mu''(\bar{\lambda}_j)}}$$
$$\times \sum_{k,s}' \frac{\sin[2\pi(k\nu_\mu(\bar{\lambda}_j) + s\bar{\lambda}_j) - \pi(k+s) + \pi/4]}{k^{3/2}}. \quad (3)$$

Here, $\nu_\mu''(\bar{\lambda}_j) = \partial^2\nu_\mu(\lambda)/\partial\lambda^2|_{\bar{\lambda}_j}$ and the $\bar{\lambda}_j$ value is determined from the relationship

$$\nu_\mu'(\bar{\lambda}_j) = \frac{\partial\nu_\mu(\lambda)}{\partial\lambda}\bigg|_{\bar{\lambda}_j} = -\frac{s}{k}, \quad 0 \leq \bar{\lambda}_j \leq \lambda_\mu. \quad (4)$$

The prime at the sum sign in Eq. (3) indicates that only the leading terms are taken into account in the sums over $s$ and $k$.

Differentiation of Eq. (3) with respect to $\mu$ gives the shell correction to the density of states,[1]

$$\Delta\rho_{sh}(\mu) = \sum_j \frac{4t_\mu(\bar{\lambda}_j)\bar{\lambda}_j}{\pi\sqrt{\nu_\mu''(\bar{\lambda}_j)}}$$
$$\times \sum_{k,s}' \frac{\cos[2\pi(k\nu_\mu(\bar{\lambda}_j) + s\bar{\lambda}_j) - \pi(k+s) + \pi/4]}{\sqrt{k}}, \quad (5)$$

while integration in Eq. (1) yields the shell correction to the free energy,

$$\Delta F_{sh}(\mu) = \sum_j \frac{\bar{\lambda}_j}{\pi t_\mu(\bar{\lambda}_j)\sqrt{\nu_\mu''(\bar{\lambda}_j)}} \sum_{k,s}' \frac{X_k^j(T)}{\sinh[X_k^j(T)]}$$
$$\times \frac{\cos\left[2\pi(k\nu_\mu(\bar{\lambda}_j) + s\bar{\lambda}_j) - \pi(k+s) + \frac{\pi}{4}\right]}{k^{5/2}}. \quad (6)$$

The following notations are introduced in these expressions:

$$X_k^j(T) = 2\pi kTt_\mu(\lambda_j), \quad t_\mu(\bar{\lambda}_j) = \pi\frac{\partial\nu_\mu(\lambda)}{\partial\mu}\bigg|_{\bar{\lambda}_j}.$$

The derivative on the left-hand side of Eq. (4) is equal to the ratio of the frequencies of the angular and radial motions of a particle with energy $\mu$ and orbital angular momentum $\bar{\lambda}_j$ [16]. The requirement that this frequency ratio be a ratio of integers is the condition for closure of the trajectory of this motion.

Equation (5) exactly coincides with the result obtained for the central potential in [17], where it was derived by using the semiclassical approximation for the Green's function. The combination of the semiclassical approach and Eq. (1) for the correction to the free energy results in a simple expression for the directly measurable quantities [Eq. (6)]. It turns out that only those electron trajectories whose energies are equal to the chemical potential of the system should be taken into account, while the condition (4) and Eq. (1), respectively, allow one to select the main trajectories and correctly include the temperature effect.

Note that the analytical approach presented in this work is much simpler and more pictorial, especially for large complexes, than the Strutinsky method of shell correction, which was applied to clusters in [14] and used earlier in nuclear physics [18, 19] and which is also based on the ETF model.

---

[1] In the semiclassical approximation, only the rapidly varying function $\sin[\ldots]$ should be differentiated.

**3.** For a fixed saddle point $\bar{\lambda}_j$, the leading contribution to the sums over $k$ and $s$ in Eq. (3) comes from the minimum values $k = \bar{k}_j$ and $s = \bar{s}_j$ that form the fraction $\bar{s}_j/\bar{k}_j$ in Eq. (4). Let us refer to the corresponding smallest length trajectory as the $j$th orbit; then the $\pi\nu_\mu(\bar{\lambda}_j)$ and $t_\mu(\bar{\lambda}_j)$ quantities have the meaning, respectively, of the radial action and the time of electron movement between the turning points along the $j$th orbit. Multiplying the numerator and denominator in $\bar{s}_j/\bar{k}_j$ by an integer $m = 1, 2, 3, \ldots$, one obtains the trajectories with $m$ periods for the movement along the $j$th orbit. One can thus replace the primed sum over $s$ and $k$ in Eqs. (3), (5), and (6) by the sum over $m$: $\sum'_{k,s}\ldots = \sum_{m=1}^{\infty}\ldots$, where $k = \bar{k}_j m$ and $s = \bar{s}_j m$.

For attractive potentials that are finite at the zero coordinate, the derivative $\nu'_\mu(0) = -1/2$ [20]. For hard self-consistent potentials, the derivative monotonically increases with $\lambda$ from $-1/2$ to $\nu'_\mu(\lambda_\mu)$ for any number of particles. For soft potentials, the situation depends on the system size; in small clusters, this derivative may decrease or be a nonmonotonic function with a minimum, while in large clusters it is an increasing function. Consequently, the sign of the second derivative $\nu''_\mu(0)$ at the zero coordinate may serve as a criterion for the potential's hardness: $\nu''_\mu(0) > 0$ for a hard potential for any cluster size.

The difference between the hard and soft potentials is illustrated in Fig. 1, where the behavior of the $\nu'_\mu(\lambda)$ derivative is demonstrated for sodium and aluminum clusters containing different numbers of electrons $N_e$. The chemical and self-consistent ETF potentials were approximated by the Saxon–Woods model potential[2]

$$U(r) = -V_0[1 + e^{(r-R)/a}]^{-1}, \quad R = r_s N_e^{1/3}, \quad (7)$$

with the parameters of the aluminum ($V_0 = 0.5319$, $a = 2.7$, $r_s = r_s^b = 2.07$, and $\mu = -0.1053$) and sodium ($V_0 = 0.22$, $a = 1.4$, $r_s = r_s^b = 3.93$, and $\mu = -0.1015$) clusters taken from [3].

Let us consider sufficiently large aluminum clusters with $N_e > 250$, for which the $\nu'_\mu(\lambda)$ derivative monotonically increases with $\lambda$ (Fig. 1). This implies that the

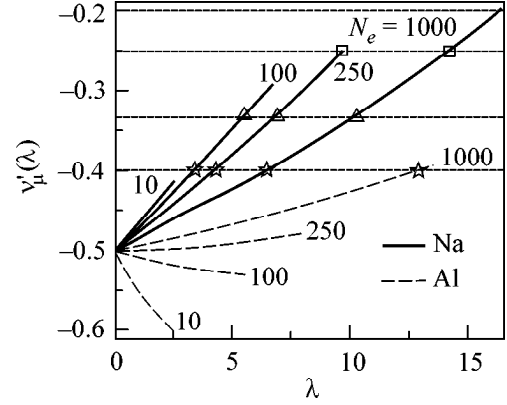

**Fig. 1.** Derivative $\nu'_\mu(\lambda)$ of the radial action with respect to the orbital angular momentum $\lambda$ for hard (Na) and soft (Al) potentials and different numbers of particles in the cluster. Calculated using potential (7) with parameters taken from [3].

rational fractions $s/k$ satisfying condition (4) lie in the range

$$-\nu'_\mu(\lambda_\mu) \leq \frac{s}{k} = \frac{\bar{s}_j m}{\bar{k}_j m} \leq \frac{1}{2}. \quad (8)$$

It then immediately follows that $\bar{s}_j \geq 1$ and $\bar{k}_j \geq 2$. The fractions with $\bar{s}_j = 1$ (1/2, 1/3, 1/4, ...) correspond to linear, triangular, square, etc. orbits. The fractions of the type $n/(2n+1)$, with $n = 2, 3, 4, \ldots$, lie between 1/3 and 1/2 and correspond to $(2n+1)$-pointed stars.

The solution $\bar{s}_j/\bar{k}_j = 1/2$ always exists and corresponds to an electron moving with zero orbital angular momentum $\bar{\lambda}_0 = 0$ along a linear orbit through the center. The corresponding contribution is small for large clusters (see [12, 13]). A circular orbit with radius $r_0$ and maximum orbital angular momentum $\lambda_\mu$ [the contribution from the upper limit of integration in Eq. (2)] is also unimportant in this case.

One can see in Fig. 1 that, for a hard sodium potential, there is a contribution from the triangular orbit ($\bar{s}_j = 1$, $\bar{k}_j = 3$) in a cluster with $N_e = 100$ and, in addition, from the square ($\bar{s}_j = 1$, $\bar{k}_j = 4$) and pentagonal ($\bar{s}_j = 1$, $\bar{k}_j = 5$) trajectories in a cluster with $N_e = 1000$. On this background, the contribution from the five-pointed star $\bar{s}_j/\bar{k}_j = 2/5$ to Eq. (6) is small due to the $\bar{\lambda}_j/\bar{k}_j^{5/2}$ factor.[3]

A completely different situation occurs for aluminum (Fig. 1). Because of the weak $\lambda$ dependence of the derivative in a soft potential, the triangular orbit appears only in very large clusters with $N_e \sim 3000$, while the smaller clusters, e.g., with $N_e = 1000$, are dominated by an orbit shaped like a five-pointed star. However, it is seen from a comparison of Eqs. (5) and (6) that the dominance of this orbit in Eq. (5) for the correction to the density of states is not as apparent as in Eq. (6); the calculations show that the oscillation amplitudes for the five-pointed and seven-pointed stars differ in Eq. (5) by only a factor of 1.36, whereas the contribution from the five-pointed star in Eq. (6) is 3.3 times greater than that from the seven-pointed star. Because of this, it is difficult to distinguish a well-defined period near the Fermi surface when considering the sum over trajectories in Eq. (5) as a function of μ at a fixed $N_e = 1000$. This might be the reason why the conclusions drawn in [10] were negative.

---

[2] The difference between potential (7) and the self-consistent ETF potential is discussed in [9].

[3] This is the reason why these secondary trajectories were not discussed in [12, 13], where only hard potentials were considered.

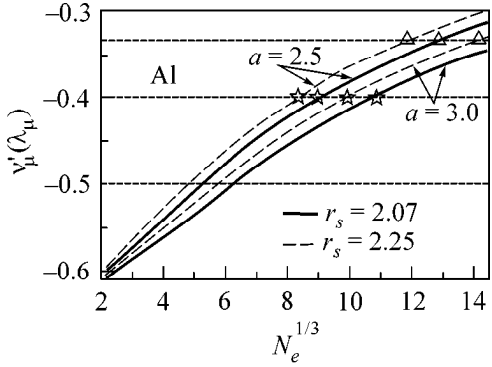

**Fig. 2.** Plots of the derivative $\nu'_\mu(\lambda_\mu)$ of radial action at the maximum orbital angular momentum vs. aluminum cluster size $N_e^{1/3}$, as calculated using potential (7) with different parameters $r_s$ and $a$.

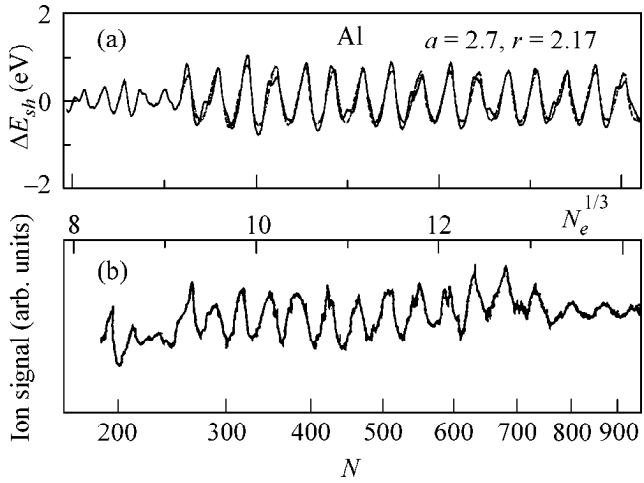

**Fig. 3.** (a) Shell correction to the total electronic energy of the aluminum cluster at $T = 0$ vs. cluster size $N_e^{1/3}$, as calculated by Eq. (6) without taking into account the triangular trajectory. (b) Oscillations in the ion signal in the mass spectra of Al$_N$ clusters [3].

**4.** It is seen from Eq. (8) and Fig. 1 that the cluster sizes $N_e^j$ for which new saddle points $\bar{\lambda}_j$ arise can be estimated from the condition $\nu'_\mu(\lambda_\mu; N_e^j) = -\bar{s}_j/\bar{k}_j$. To do this, it is sufficient to know the dependence of $\nu'_\mu(\lambda_\mu)$ on the cluster size $N_e$. The relevant function calculated for aluminum by the formula $\nu'_\mu(\lambda_\mu) = [3 + U''(r_0)r_0^2/p_\mu^2(r_0)]^{-1/2}$ [13] is shown in Fig. 2 for various values of parameters $r_s$ and $a$. One can see in Fig. 2 that a decrease in $r_s$ is equivalent to an increase in $a$, i.e., to the softening of the potential.

Self-consistent calculations suggest [9] that $r_s$ depends weakly on $N_e$. Hence, as $N_e$ increases, the curves corresponding to large $r_s$ should gradually transform into a curve corresponding to the radius $r_s = r_s^b$ of the Wigner–Zeits cell in bulk metal. To account for this effect, the calculations were carried out first for the shell correction (6) to the total electronic energy of Al$_N$ at $T = 0$ using potential (7) with $r_s = 2.17 > r_s^b$. Figure 3 presents (a) the results of calculations without inclusion of the triangular trajectories and (b) the measured oscillations in the mass spectra of aluminum clusters. The dashed line in the calculated curve (Fig. 3a) indicates the contribution from the five-point-star trajectory. It dominates at $N_e > 250$ ($N_e^{1/3} > 9.1$), while the contributions from other star-shaped trajectories are small in this mass range. The cosine argument for the five-point-star trajectory depends linearly on $N_e^{1/3}$:

$$5\nu_\mu(\bar{\lambda}_*) + 2\bar{\lambda}_* = -2.75317 + 3.1487 N_e^{1/3}, \quad (9)$$

resulting in a periodic dependence on this variable. At $N_e > 250$, curve (a) fits curve (b) well, both structurally and in period. Note also that the chaotic portions of the calculated and experimental spectra exhibit similar behavior at $N_e < 250$ (calculations show that it is caused by the contributions from the star-shaped trajectories with $\bar{s}_j/\bar{k}_j = 3/7, 4/9, 5/11$, and $6/13$).

A triangular trajectory with $r_s = 2.17$ is expected to appear at $N_e^\Delta = 2450$ ($N_e^{1/3} \simeq 13.5$). Since the corresponding oscillation amplitude is ~3.3 eV, this trajectory immediately becomes dominant; it oscillates with almost halved frequency. The relevant term in the cosine argument depends linearly on $N_e^{1/3}$: $3\nu_\mu(\bar{\lambda}_\Delta) + \bar{\lambda}_\Delta = -3.2568 + 1.7605 N_e^{1/3}$. However, when considering the above-mentioned $r_s(N_e)$ dependence, one should use a smaller $r_s$ value for the self-consistent potential in approximation (7) in this range of $N_e$ values. This is indirectly confirmed by a small (although distinguishable in Fig. 3b) change in the frequency of experimental oscillations at $N_e^{1/3} > 13$. We note parenthetically that the frequency would not change if the spectrum were determined by the ion lattice.

Setting $r_s = r_s^b = 2.07$, one obtains for the five-pointed star the $5\nu_\mu(\bar{\lambda}_*) + 2\bar{\lambda}_* = -2.9435 + 3.0249 N_e^{1/3}$ dependence instead of Eq. (9); i.e., the period indeed increases slightly. The $N_e^\Delta$ value for this $r_s$ value is equal to 2750 ($N_e^{1/3} \geq 14$), and the spectra should be rearranged at $N > 900$.

To estimate the temperature-induced decay of the electronic shell oscillations, the temperature multiplier in Eq. (6) should be evaluated for the five-pointed star. The calculations show that the characteristic reciprocal temperature $2\pi\bar{k}_j t_\mu(\lambda_j)$ increases for this orbit from 900 to 1350 upon increasing $N_e$ from 750 to 3000. It follows that the temperature multiplier differs (for $m = 1$) only slightly from unity at $T = 300$ K $\simeq 0.001$ au and decreases, respectively, from 0.88 to 0.75. For the triangular trajectory, the characteristic reciprocal temperature is on the order of 600 at $N_e \sim 3000$ and the corresponding multiplier is 0.95. Therefore, if the lattice is molten at $T = 300$ K, the oscillations due to electronic shells should manifest themselves in full measure.

This work was supported in part by the Russian Foundation for Basic Research, project no. 00-01-00397.

*Translated by V. Sakun*